# XFLEX HYDRO demonstrators grid services assessment and Ancillary Services Matrix elaboration

**C. Nicolet, M. Dreyer, C. Landry**
**S. Alligné, A. Béguin**
Power Vision Engineering Sàrl
Rue des Jordils 40
CH-1025 St-Sulpice
Switzerland

**Y. Vaillant, S. Tobler, G. Sari**
GE Renewable Energy
Zentralstrasse 40
CH-5242 Birr
Switzerland

**G. Païs**
CEA, LITEN, INES
Univ. Grenoble Alpes
73375 Le Bourget-du-lac
France

**M. Bianciotto, S. Sawyer, R. Taylor**
International Hydropower Association
One Canada Square
E14 5AA London
United Kingdom

**M. V. Castro, M. H. Vasconcelos, C. Moreira**
INESC TEC
Campus da FEUP, Rua Dr Roberto Frias
4200-465 Porto
Portugal

**Intro**

The overarching objective of the Horizon 2020 European Project XFLEX HYDRO [1] is to develop and demonstrate new technological solutions capable of being integrated in different types of Hydro Power Plants (HPP) aiming to improve their efficiency and performances regarding the provision of several Electric Power Systems (EPS) services. Such developments are expected to actively contribute to the decarbonization of the European EPS, allowing HPP to have increased capabilities for the provision of advanced grid services in face of an evolving EPS characterized by increased shares of time variable renewable energy sources, [2].

The activities developed within the scope of Work Package 2 of the XFLEX HYDRO project allowed to perform a detailed assessment of current and emerging flexibility services of EPS and their associated market frameworks. The study focused in particular on the countries where XFLEX HYDRO demonstrations are taking place (Portugal, France and Switzerland). In summary, the basket of ancillary services under evaluation is composed as follows: Synchronous inertia, Synthetic inertia, Fast Frequency Response (FFR), Frequency Containment Reserve (FCR), Automatic Frequency Restoration Reserve (aFRR), Manual Frequency Restoration Reserve (mFRR), Replacement Reserve (RR), Voltage/reactive power control (Volt/var) and Black start capability.

This paper presents the methodology and key results which enabled to establish the so-called Ancillary Service Matrix (ASM) presenting the ability to deliver the different ancillary services of each of the 6 demonstrators of the project combined with the applicable technologies studied in this analysis. These technologies include i) the variable speed technology with Doubly Fed Induction Machine (DFIM) or Full Size Frequency Converters (FSFC), ii) the Smart Power Plant Supervisor (SPPS) enabling to extend the operating range of the hydraulic units in turbine mode based on a better knowledge of the hydro unit wear and tear and associated costs over the full unit operating range, iii) the hydraulic short circuit (HSC) operation leading to simultaneous operation of pump and turbines of Pumped Storage Power Plants (PSPP) and iv) the Hydro-Battery-Hybrid (HBH) applied at Run-of-River demonstrator.

The demonstrators considered for this study includes i) 4 pumped storage power plants: Grand Maison (FR), Frades 2 (PT), Alqueva II (PT) and Z'Mutt (CH), as well as ii) 1 conventional hydro storage plant: Alto Lindoso (PT) and iii) 1 run-of-the-river plant: Vogelgrün (FR), see short demonstrator overview in Annex 1 and [1]. For each demonstrator a 1D SIMSEN simulation model was developed and validated and was further enhanced to include the model of control system enabling to address the various ancillary services. The systematic 1D numerical simulation of ancillary service contribution of each demonstrator and related technologies enabled to quantify the magnitude of active power response to contribute to the different grid services. Overall, more than 130 simulations have been performed to quantify all ancillary services. The results have been scored between 0 and 5 for each ancillary service, allowing to populate the Ancillary Service Matrix which is summarizing the results in a graphical and synthetic way. The analysis of the score of the Ancillary Services Matrix enables the reader to draw several key conclusions about the benefits unlocked by the implementation of these technologies which are summarized in the paper.

## 1. Objectives of the Ancillary Services Matrix elaboration

One of the key objectives of the Work Package 2 of XFLEX HYDRO was to evaluate the performances of different hydroelectric technologies and solutions applied to the XFLEX HYDRO demonstrators by means of 1D system simulations for each set of flexibility products which have been identified as relevant for the future European EPS stability, see [3]. Therefore, time domain simulations have been performed for each demonstrator and associated relevant technologies to quantify the magnitude of the different flexibility products. The results are converted into scores which are then used to populate the so-called Ancillary Service Matrix (ASM), see Figure 1.

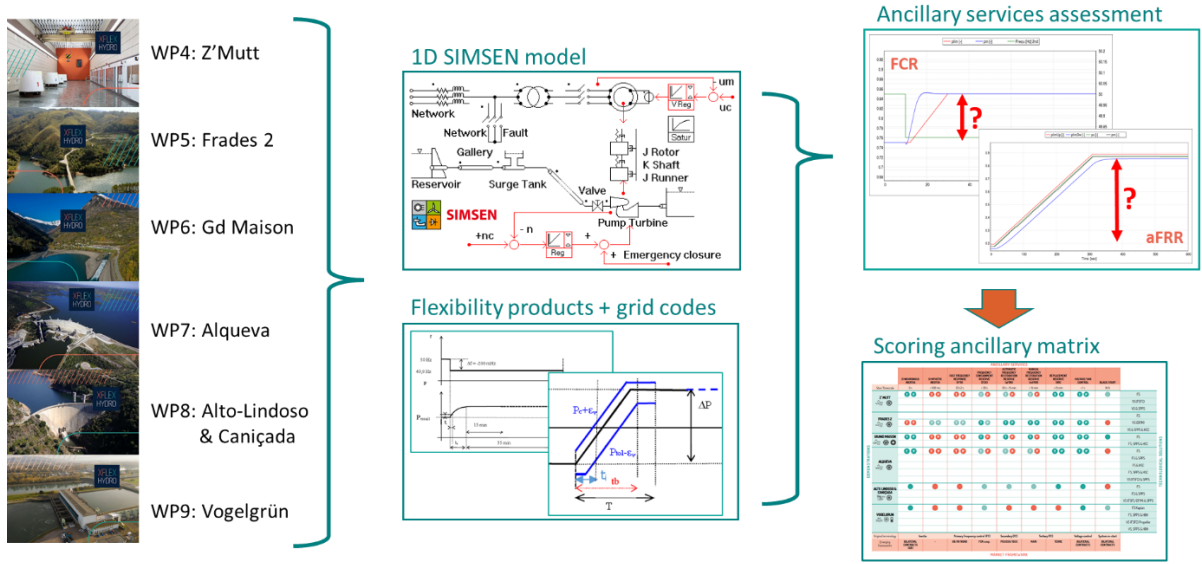

*Figure 1 Methodology used to construct the Ancillary Services Matrix.*

The technological solutions to be assessed and/or implemented within the scope of the XFLEX HYDRO project are as follows:
- **Fixed speed (FS)** units;
- **Variable speed (VS)**, through the installation/conversion to Doubly Fed Induction Machine (DFIM) or Full-Size Frequency Converter (FSFC);
- **Smart Power Plant Supervisor (SPPS)**, a methodology to provide an adequate monitoring and extensive knowledge of the machine that will enable the plant owner to manage the risks and costs associated to a temporary off-design operation according to the potential benefits it offers. In this project, SPPS methodology is considered as an umbrella of all the actions, studies and control measures performed to increase the controllability and monitoring of the HPP without the installation of significant hardware;
- **Hydraulic Short Circuit (HSC)**, which allows the hydro power plant (HPP) to operate at the same time a turbine in parallel to a pump. In case of fixed speed units, this enables the simultaneous operation of a pump at constant power while performing power control with the turbine, varying continuously the consumption of the plant. When variable speed is considered, the operating range of the pump and of the turbine can be combined to further extend the overall operating range;
- **Hydro-Battery-Hybrid (HBH)**, namely the hybridization of the HPP through the installation of a Battery Energy Storage System (BESS).

The breakdown of the technological solutions per demo portrayed in the matrix follows a preliminary assessment considering demo owners and associated equipment manufacturers view regarding the envisioned development plan for each power plant, either in terms of engineering/simulation studies, reduced scale models tests and on-site live tests.

For each demo, the set of technological solutions to be assessed by means of numerical simulation is compared to the baseline condition corresponding to the existing configuration of the given demonstrator which are the following:

- **Z'Mutt:**
    - Fixed speed;
    - Fixed speed with SPPS;
    - Variable speed with a FSFC (baseline);
    - Variable speed and SPPS.
- **Frades 2:**
    - Fixed speed;
    - Fixed speed with SPPS;
    - Fixed speed with SPPS and HSC;
    - Variable speed with a DFIM (baseline);
    - Variable speed with SPPS;
    - Variable speed, SPPS and HSC.
- **Grand Maison:**
    - Fixed speed (baseline);
    - Fixed speed, SPPS and HSC.
- **Alqueva II:**
    - Fixed speed (baseline);
    - Fixed speed with SPPS;
    - Fixed speed with SPPS and HSC;
    - Variable speed with a FSFC;
    - Variable speed with SPPS;
    - Variable speed, SPPS and HSC.
- **Alto Lindoso:**
    - Fixed speed (baseline);
    - Fixed speed and SPPS;
    - Variable speed with a FSFC;
    - Variable speed and SPPS.
- **Vogelgrün:**
    - Fixed speed (baseline);
    - Fixed speed, SPPS and HBH;
    - Variable speed (FSFC).

## 2. Methodology to develop the Ancillary Services Matrix

### 2.1. General approach

The engineering work supporting the assessment of the different HPP technologies with respect to the basket of identified ancillary services relies on 1D SIMSEN simulation models that have been developed and validated in the framework of each project demonstrator initial studies, [4], [5], [6]. Afterwards, these models were enhanced to include the model of control system related to both fixed speed and variable speed technologies when applicable. The control structure considered for variable speed technology allows to provide Frequency Containment Reserve (FCR), Synthetic Inertia (SI), Fast Frequency Response (FFR) and automatic Frequency Restauration Reserve (aFRR) ancillary services, see [5]. The control structure considered for fixed speed includes in principle an active power and speed control loops which are combined with permanent droop enabling to address FCR and aFRR ancillary services, [4]. Additionally, for Vogelgrün, the 1D SIMSEN simulation model of the Kaplan turbine and related turbine governor was coupled with a MATLAB model of the Battery Energy Storage System and related HBH joint control, using Functional Mockup Interface (FMI), [4].

The evaluation of the different ancillary services required to select grid codes, whenever applicable, or to define relevant metrics to quantify the contribution of the demonstrators and related technologies to the different ancillary services. The French Grid Code of RTE has been selected for the evaluation of FCR and aFRR capabilities [7], while ENTSO-E grid code for Nordic Synchronous Area was considered to evaluate FFR capability, [8]. It was assumed that the contribution to aFRR is also representative of the possible contribution to mFRR and RR services. For Voltage/Var control, the magnitude of reactive power that a hydro unit can deliver at maximum active power was considered to represent the contribution to this ancillary service, without need to perform numerical simulations. For black start capability, time domain simulations were performed to quantify the maximum isolated resistive load that could be directly connected to a hydro unit operating at speed no load at 50 Hz, and ensuring that the isolated grid frequency would not fall below 49 Hz when fixed speed technology was considered. When variable speed technology was considered, the full range of speed control was considered for DFIM technology to maximize the contribution to black start, while the limit for FSFC technology was the minimum rotational speed before the unit would stall and result in abrupt shutdown. It is important to mention that variable speed units need to be able to operate in grid forming mode to perform black-start, [9], [10]. Moreover, black start capability also requires some external source of energy (battery, auxiliary hydro unit, diesel, etc) and voltage/reactive power regulator shall enable electrical line energization and proper voltage control. All the simulations have been performed under minimum head condition which corresponds to the most detrimental operating condition regarding the provision of ancillary services.

Finally, the evaluation of synchronous and synthetic inertia contributions required a special attention since there is no grid code nor metrics available for these ancillary services. An analytical expression was derived to quantify the magnitude of the active power that a synchronous machine would inject or absorb in case a large power network frequency deviation would occur, [11]. This simple analytical expression, which was validated with detailed time domain simulations, shows that the magnitude of active power injected or absorbed is proportional to the hydro unit mechanical time constant $\tau_m$ for a given Rate of Change of the frequency (*RoCoF*). A typical *RoCoF* value of 1 Hz/s during 1 s was considered to quantify active power injection or absorption. The analytical approach also shows that units operating in turbine or in pump mode both contribute to synchronous inertia, since each feature rotating masses storing angular kinetic energy in the same way as a flywheel.

Unlike synchronous machines, the rotational speed of variable speed units is decoupled from the power network frequency. Therefore, the inertial response of a variable speed unit must be emulated using an appropriate control structure to provide so-called synthetic inertia to the power network. Inertia emulation control structure is based on the swing equation to achieve active power injection or absorption which is also proportional to the *RoCoF*, like for a synchronous machine, and thus replicate the corresponding flywheel effect. Time domain simulations performed enabled to show that variable speed units inject or absorb the same active power as a synchronous machine in the event of a frequency variation, provided that the power network features enough synchronous inertia so that the variable speed unit can be operated in grid following mode, [11].

The flexibility products to be evaluated are the following:
- Synchronous inertia;
- Synthetic inertia with response time < 0.5 s;
- Fast Frequency Response (FFR) with response time < 0.5 s – 2 s;
- Frequency Containment Reserve (FCR) with response time < 30 s;
- Automatic Frequency Restoration Reserve (aFRR) with response time < 30 s – 5 min;
- Manual Frequency Restoration Reserve (mFRR) achievable in less than 15 min, with a maximum full activation time of 12.5 min;
- Replacement Reserve (RR) to be made available in less than 30 min;
- Voltage/Var control with response time < 1 s;
- Black-Start capability.

The 1D SIMSEN models of the demonstrators developed in the framework of the initial studies are displayed in Figure 2. These models include detailed representation of the hydropower plants 1D dynamic behaviour including water hammer phenomena in waterways, mass oscillations in surge tanks, 2 or 4 quadrants quasi-static characteristics of the hydraulic machines and unit rotating inertia effects, [12]. The detailed modelling of the relevant control system of the different technologies were also included in the 1D models. Special care has been paid to develop DFIM and FSFC variable speed control modelling which also included inertia emulation and FCR control capabilities. This control structure can be used to provide synthetic inertia, FFR and FCR services with variable speed units, [5]. The modelling of HBH control applied to the Vogelgrün demonstrator also required particular attention to take into account the joint control algorithm that manages hydraulic turbine and BESS active power to achieve FCR control service while minimizing wear and tear of the Kaplan turbine blade pitch angle positioning mechanism, [13]. The different technologies addressed by means of 1D system simulation for all the demonstrators are present in Table 1, with the indication of the related contributors. Then, the evaluation of the different flexibility products required the selection of relevant grid code or the definition of appropriate metrics when no grid code specifications were available. A proper definition of the simulation scenarios and performance criteria to score the different ancillary services was elaborated for each flexibility product. Based on these scenario definitions, systematic 1D system simulations of the different flexibility services were performed for each demonstrator and corresponding performances were quantified and converted into scores ranging between 0 and 5 to populate the Ancillary Service Matrix. The score 0 means no capability, and the score 5 means maximum capability according to the service associated metric.

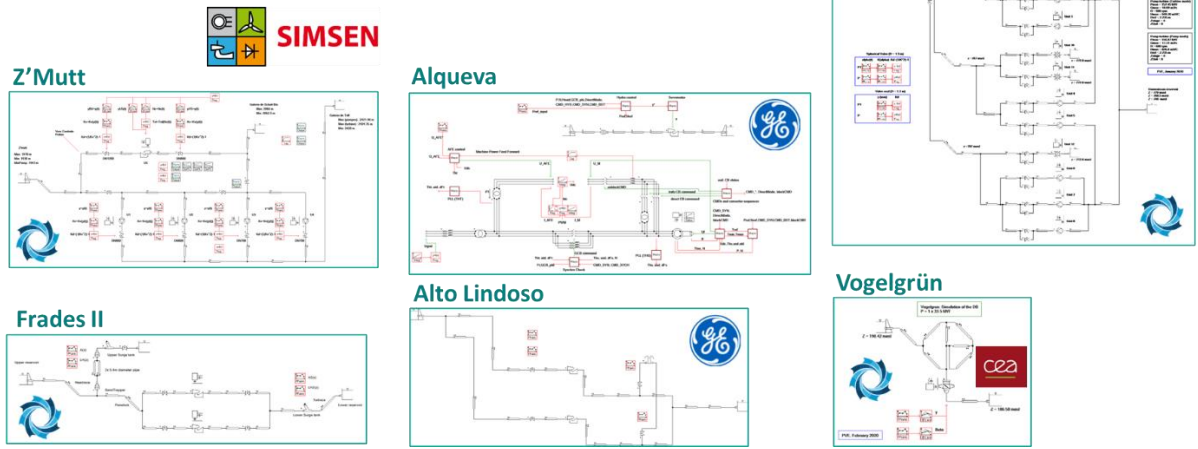

*Figure 2  1D SIMSEN models of the XFLEX HYDRO demonstrators.*

*Table 1  Overview of the 1D modelling of the demonstrators and the different technologies which have been modelled and simulated. HSC: Hydraulic Short Circuit, NA: Not Applicable, BESS: Battery Energy Storage System, ☑: Applicable.*

| Demos | Resp. | Fixed speed control and parameters | Variable speed control and parameters | BESS hybrid control | Control in HSC |
|---|---|---|---|---|---|
| WP4: Z'Mutt | PVE | ☑ | ☑ | NA | NA |
| WP5: Frades 2 | PVE | ☑ | ☑ | NA | ☑ |
| WP6: Gd Maison | PVE | ☑ | NA | NA | ☑ |
| WP7: Alqueva II | GE HYDRO | ☑ | ☑ | NA | ☑ |
| WP8: Alto-Lindoso | GE HYDRO | ☑ | ☑ | NA | ☑ |
| WP9: Vogelgrün | PVE and CEA | ☑ | ☑ | ☑ | NA |

## 2.2. Ancillary services description

The work developed within Work Package WP2 provided a detailed review regarding the basket of the ancillary services being required by grid operators in European synchronous areas. Additional details of the specific characteristics of those services are presented in Deliverable D2.1 [3]. The review of the different network services has enabled the selection of those that are expected to have a relevant impact on the XFLEX HYDRO project, taking into account the different characteristics of the demonstrators and associated technological options, being briefly summarised hereafter:

- **Synchronous inertia:** This is the inherent capability of rotating machinery directly connected to the power grid (loads included) to store kinetic energy in the rotating mass of the rotor and inject this kinetic energy to the grid. This has a fundamental role in supporting the grid frequency transient behaviour in the moments subsequent to an active power imbalance. The synchronous inertia has a major impact in the rate at which the grid frequency varies – Rate of Change of Frequency – *RoCoF* (Hz/s) following a system incident. The grid frequency transients are of major importance: as a result of system inertia reduction, it risks the activation of underfrequency load shedding and generators tripping following a system incident. Although load shedding actions contribute to mitigate frequency transients, it is a last resource/defence plan at the disposal of the system operator which should be avoided (as a measure of the system security/robustness, the activation of underfrequency load shedding represents a loss of system dynamic security). Sudden generation tripping as a result of large *RoCoF* further aggravates the grid frequency transients. Synchronous inertia may be obtained by increasing the rotating mass connected to the system, which includes the one in synchronous generators, pumps of pump-storage HPPs with fixed speed units, as well as throughout the synchronous condenser operation mode that can be exploited in pumped storage plants. Lack of synchronous inertia and the definition of minimum volumes of synchronous inertia in the power grids is currently a key concern of modern power systems with increased shares of converter-interfaced generation units.

- **Synthetic inertia:** the Commission Regulation 2016/631 of 14 April 2016, establishing a network code on requirements for grid connection of generators, defines 'synthetic inertia' as the capability provided by a power park module to replace the effect of inertia of a synchronous power-generating module to a prescribed level of performance. The primary benefit/need of synthetic inertia provision is related to the need of limiting the initial *RoCoF* of low synchronous inertia systems, following a worst-case reference disturbance, to keep it below the maximum withstanding capability of connected grid users (either demand and power generation units) to remain connected.

  In general, converter interfaced energy sources can provide short-term frequency support through proper control of the coupling interface. While not providing synchronous inertia, they are able to swiftly adapt active power output, driven by their control system, to deliver "synthetic inertia", provided some energy buffer is available within the primary energy source as it is the case of the kinetic energy stored in the rotating masses of wind or hydro facilities. Recent studies, performed within the EU project Eu-Sysflex [14], demonstrated that, for the Continental Europe EPS, the increasing integration of converter interfaces for renewable energy sources such as wind and solar tends risks the increase in the *RoCoF* in particular areas of the system. The studies conducted evidence that the estimated *RoCoF* values are generally below 1 Hz/s for most Continental Europe countries following the reference incident (-3 GW in France, related to the tripping of the 2 largest nuclear plants, and -2 GW in the other zones). However, when the Iberian Peninsula is considered, the low synchronous interconnection capacity with France and the increasing share of converter interfaced generation points towards *RoCoF* values up to 1.3 Hz/s, hence demonstrating a more localised scarcity emerging for this specific service.

- **Fast Frequency Response (FFR):** FFR is usually designed to provide an active power response faster than existing operating reserves, typically in less than 2 seconds, in the timeframe following inertial response (i.e., typically after 500 ms) and before activation of the frequency containment reserve service (which has a maximum delay of 2 seconds). Also emerging from reducing system inertia and increased *RoCoF*, it is verified that conventional frequency containment reserves start to have a more constrained operation time windows to have an effective response to limit the frequency nadir/zenith to the prescribed range in each area. Recent studies, performed within the EU project Eu-Sysflex [14], demonstrated that, for the Continental Europe case, frequency nadir following the loss of a large generating unit tend to decrease as penetration levels of converter-interfaced generation in the corresponding area increase. Nevertheless, the frequency nadirs are still above the threshold for load shedding (49.2 Hz). The particular case of Iberian Peninsula reveals some concerns, risking a frequency nadir of 49.35 Hz for a 2 GW loss in the corresponding area, as a result of the low synchronous interconnection capacity with central Europe and low regional inertia due to the high penetration of non-synchronous generation. FFR is already being explored in European areas such as in United Kingdom (previously tendered as Enhanced Frequency Response), the Irish and the Nordic markets with different specifications associated to this product. However, a common characteristic can be clearly identified in the areas where this product is being required: the time that takes to fully activate this product should be less than 2 seconds.

- **Frequency Containment Reserve (FCR)**, which, in former terminology, is known as primary frequency control capability of generation units: FCR aims to contain system frequency after the occurrence of an active power imbalance, by maintaining the balance between active power generation and demand within a synchronous area and aiming to comply with pre-defined frequency metrics. For FCR providers in the Continental Europe synchronous area, the service must be fully activated within 30 seconds and the power-generating module shall be capable of providing full active power-frequency response for a period between 15 to 30 minutes (specified by the Transmission System Operators (TSOs) of each synchronous area). The minimum technical requirements to be ensured by FCR providers is presented and discussed in detail in Deliverable D2.1 [2].

- **Automatic Frequency Restoration Reserve (aFRR)** which, in former terminology, is known as secondary frequency control: aFRR is an automatic process aiming to restore the system frequency back to its set point (normal) value and to keep the power interchange program among Load-Frequency Control (LFC) areas. The required volumes of this service are managed by TSOs under specific market arrangements. Regarding system operation, the activation of the service is controlled centrally by TSOs. In Continental Europe, this service already has a standard product, properly defined to be traded in a European platform for the exchange of balancing energy from aFRR. This standard product is developed in close coordination with the PICASSO initiative, a project established by European TSOs for the implementation of an aFRR platform. According to EU regulation, the aFRR activation delay must not exceed 30 s. Besides, a maximum of 5 min is set for the Full Activation Time (FAT) of the aFRR standard product (starting from December 2025).

- **Manual Frequency Restoration Reserve (mFRR)** which, in former terminology, is known as directly activated tertiary frequency control: The provision of mFRR is a frequency restoration process, having therefore similar goals to aFRR. It is activated by means of TSO instructions for manual FRR deployment in the associated LFC area. This service has a standard product defined to be exchanged in a European platform for the exchange of balancing energy from mFRR, which is specified in close coordination with the MARI project [15]. According to EU regulation, mFRR must be fully activated in 15 minutes. Within the mFRR standard product, the FAT is set to a maximum of 12.5 minutes (to consider the delay between an event and the manual transmission of reserve activation orders) and has a minimum delivery period of 5 minutes.

- **Replacement Reserve (RR),** which, in former terminology, is known as scheduled activated tertiary frequency control: The aim of the RR process is to progressively replace and/or support the frequency restoration control process in the disturbed control area. RR implementation consists in optional TSO instructions for manual activation of reserves in the LFC area, usually performed after the time to restore frequency, in a timeframe between 15 minutes up to 1 h. According to EU regulation, defining the minimum technical requirements of RR providers is a responsibility of the TSOs of each LFC block. This service also has a standard product to be exchanged in a European platform for the exchange of balancing energy from RR, which is specified in close coordination with the TERRE project [16]. The main characteristics for a RR standard product include a FAT of 30 min and a minimum delivery period of 15 minutes.

- **Voltage/reactive power control (Volt/var):** The Volt/var control process is implemented by manual or automatic control actions, being intended to control network node voltages within specific ranges. The requirements for this control are highly dependent of local or regional characteristics of the power system. Nowadays, this is typically a mandatory and not remunerated service. In some countries, there are also bilateral contracts held between service providers and the TSO for the provision of extra Volt/var control, which is remunerated according to each specific contract. Often, upon request by the local TSO, pump storage HPP operate in the synchronous condenser mode to provide Volt/var support to the grid.

- **Black-Start:** Black-Start is the process of autonomous restarting operation of a power plant during a grid blackout, from a completely non-energized operating state and without any power feed from the network. This type of service is intended to re-energize network sections while providing cracking power to other plants to support its restart, as well as to progressively supply system loads and bring the grid system back to normal operating conditions. Fundamental characteristics of Black-Start capable power plants are related to the autonomous re-start without external grid support, capability of energizing unloaded transmission lines, capability of picking up load blocks (amounts depending on the plant type), hence requiring the capability of assuring a proper power-frequency regulation. Black-Start is nowadays contracted through bilateral agreements, and it is not expected to have its own market framework in the future. Nevertheless, HPPs are particularly well suited for Black-Start and it is expected that, in future scenarios with a massive increase of renewable power production, the Black-Start capabilities of HPPs will become of crucial importance to guard system security.

### 2.3. Ancillary services evaluation criteria

#### 2.3.1. FCR ancillary service evaluation criteria (RTE)

This ancillary service aims to contain system frequency after the occurrence of an active power imbalance, by maintaining the balance between active power generation and demand. Experimentally, the qualification test imposes a step frequency deviation from the nominal value of -200 mHz and +200 mHz to the turbine controller as input. The output power response of the unit (Primary Control Reserve, *RP*) must respect the limitations defined in the Grid codes. For a frequency variation of -200 mHz, the RTE Grid code [7] has been selected as the reference document and the limitations are illustrated in Figure 3, being defined by the following parameters:

- *ev*: Measurement uncertainty. $ev = 5\%$ of primary reserve $R_P$.

- $t_i$ : Time after which the power response is greater than the measurement uncertainty. This time must be less than 2 s.

- $t_r$ : Time at which the power response reaches 100% of the primary reserve $R_P$. This time must be less than 30 s.

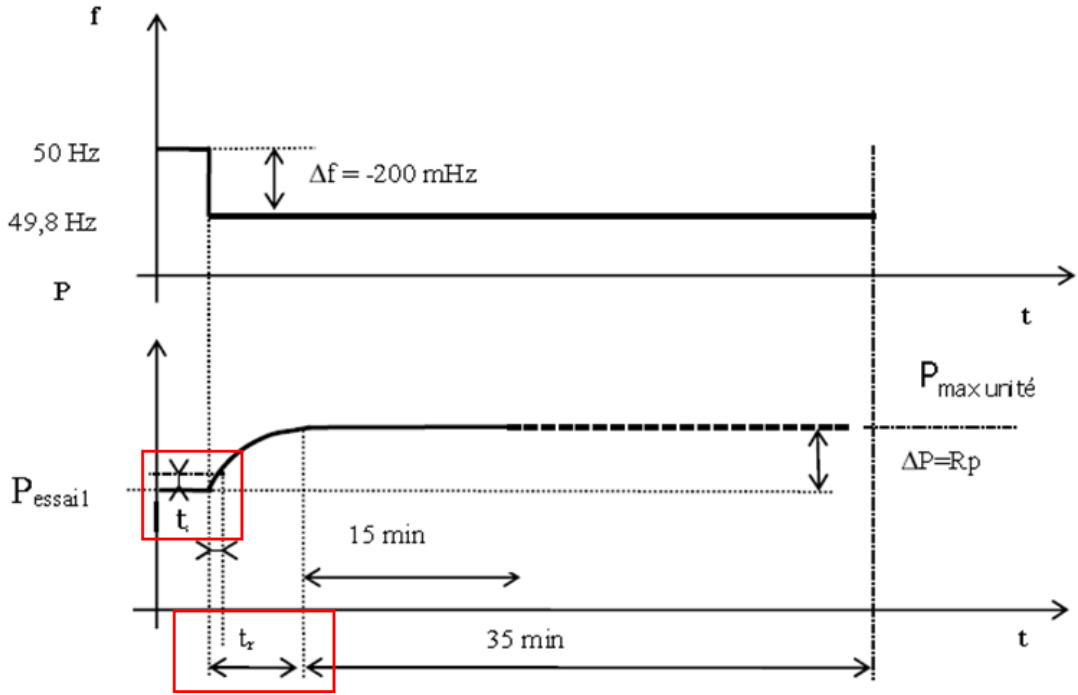

*Figure 3  RTE definition of limitation for the output power response after a frequency variation of -200 mHz [7].*

### 2.3.2. aFRR ancillary service evaluation criteria (RTE)

This ancillary service is an automatic process aiming to restore the system frequency back to its set point value and to keep the power interchange program among load-frequency control (LFC) areas. Experimentally, to validate the qualification test, the power set point varies with an amplitude of *2·PR* in a time of 300 s. The minimum power set point is defined for low head operating conditions *Hmin*. The power limitations to be respected are defined by the RTE grid code [7] and the different parameters are described below:

- *ev* : Measurement uncertainty. *ev* = 5% of secondary reserve *PR*.
- $T_b$: Response time after which the secondary reserve band is released.
- $T_i$ : Time after which the power response is greater than the measurement uncertainty ev. $T_i$ < 2 s.
- *T* : Ramp time (300s) increased by 100 s.
- $P_c$: Power set point.
- *PR*: Secondary reserve band.
- $P_{tol}$: Filtering of the setpoint by a time constant of 20 s.

<u>For the loading case</u>, to pass the qualification test, the black curve corresponding to the output power should be located between upper and lower limits (blue lines), see Figure 4.

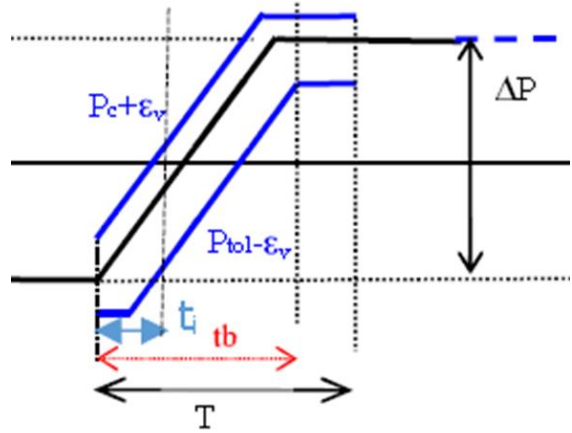

*Figure 4  RTE power limitation definition for the aFRR (loading) [7].*

### 2.3.3. FFR ancillary service evaluation criteria (RTE)

FFR is designed to provide an active power response faster than existing operating reserves, typically in less than 2 seconds, in the timeframe following inertial response (i.e., typically after 500 ms) and before activation of FCR (which has a maximum delay of 2 s).

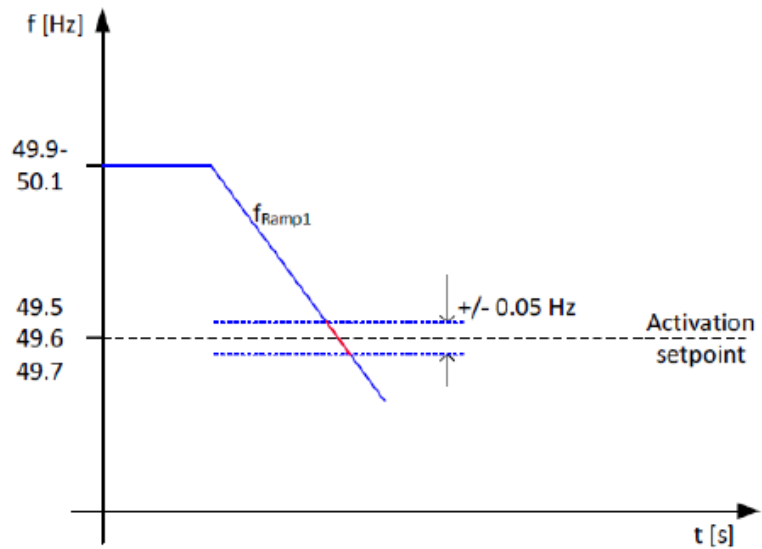

*Figure 5  Ramp test for FFR compliance verification.*

For the qualification test, this ancillary service is activated by a frequency drop according to a step or a ramp, see Figure 5. This *activation instant* is the reference time for this test and is fixed to zero. According to the ENTSO-E Grid Code for Nordic Countries selected for the FFR evaluation [8], the *maximum full activation time* must respect one of the following 3 alternatives according to the *activation level*:

- The maximum time for full activation is 0.70 s for the activation level 49.5 Hz.

- The maximum time for full activation is 1.00 s for the activation level 49.6 Hz.

- The maximum time for full activation is 1.30 s for the activation level 49.7 Hz.

The provider may choose any of the three alternatives, but the choice must be specified beforehand. In this document, the third proposition was selected, maximum activation time = 1.30 s. During the *activation time*, the unit must deliver a power amplitude, named *FFR capacity*, and maintains it during the *minimum support duration* 5.0 s (for short support duration) or 30 s, for long support duration, see Figure 6. A *maximum acceptable over delivery* is fixed to 20% of the prequalified FFR capacity. However, the reserve connecting TSO may allow up to 35% over delivery upon request, depending on the national procurement process. Irrespectively, the duration of the FFR providing entities full FFR provision cycle must be shorter than 15 minutes, to be ready for a new activation. Finally, there is no limitation on the rate of deactivation for the long support duration FFR, i.e., the deactivation can be stepwise.

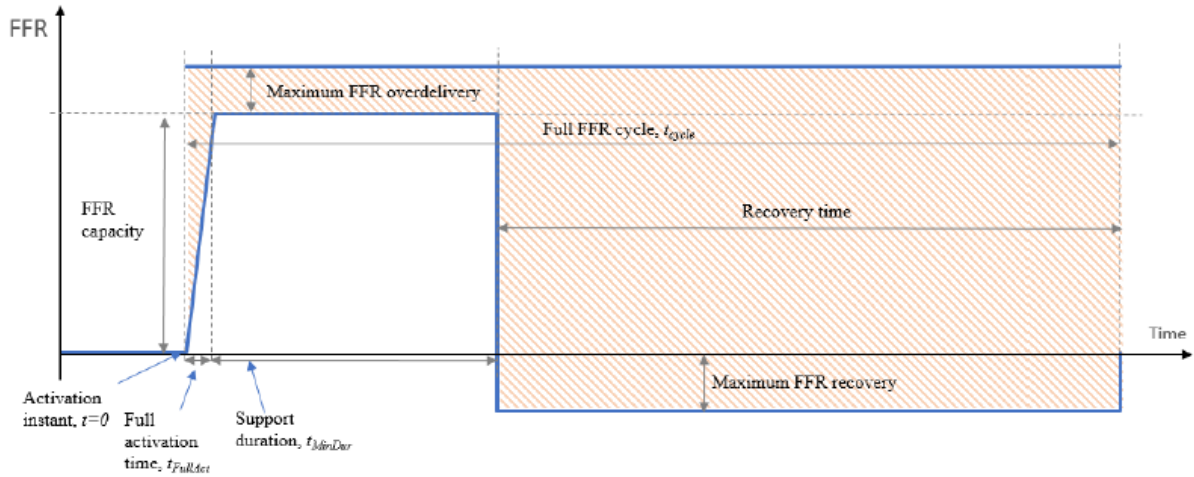

*Figure 6 FFR recovery requirement for long duration FFR in the ENTSO-E document [8].*

## 3. Example of Frades 2 ancillary services evaluation

The key simulation results obtained for the demonstrator of Frades 2 are presented in the following sub-chapters in order to illustrate the methodology adopted to evaluate the ancillary services of the 6 different XFLEX HYDRO demonstrators.

### 3.1. Power Plant 1D SIMSEN modelling

Frades 2 hydroelectric plant is a PSPP built between 2010 and 2017 on the Rabagão river in the north of Portugal. The plant is composed of two high head, variable speed units made of two reversible pump turbines, coupled with 420 MVA DFIM which are currently the Europe largest and most powerful machines. The main hydraulic and power generation characteristics are given in table 2, [17]. The waterway includes a headrace tunnel, an upper surge tank followed by a sandtrap, a penstock that feeds the distributor of each unit, a lower surge tank and the tailrace tunnel. The reversible pump-turbine is characterized by a specific speed of $N_q=38$ SI and a unit mechanical time constant of $\tau_m = 7.9$ s. In the framework of the XFLEX HYDRO project, it is planned to perform tests to demonstrate the possibility to operate the pumped storage power plant with hydraulic short-circuit operation as illustrated in Figure 7. To quantify the capacity of ancillary services, the allowed operating range of the Frades 2 units must be considered as boundary condition for the simulations. In particular, with a minimum head, the normal operating range of the turbine mode varies from $P_{min} = 186.4$ MW to $P_{max} = 372.8$ MW, while in pump mode the input power can be varied from $P_{min} = -300$ MW to $P_{max} = -390$ MW.

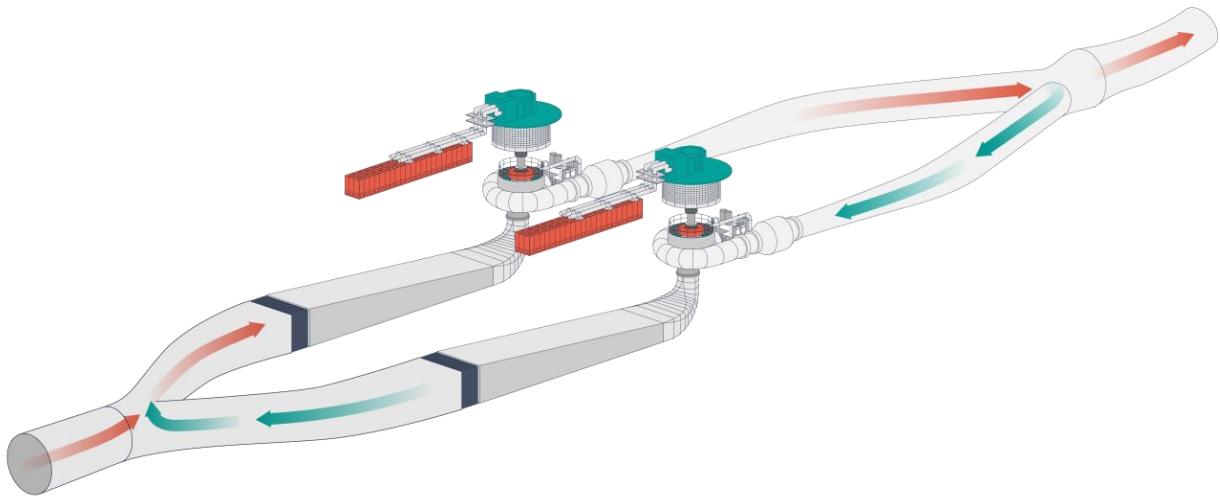

*Figure 7 Schematic representation of the two DFIM variable speed reversible pump-turbine units of Frades 2 operated in hydraulic short circuit.*

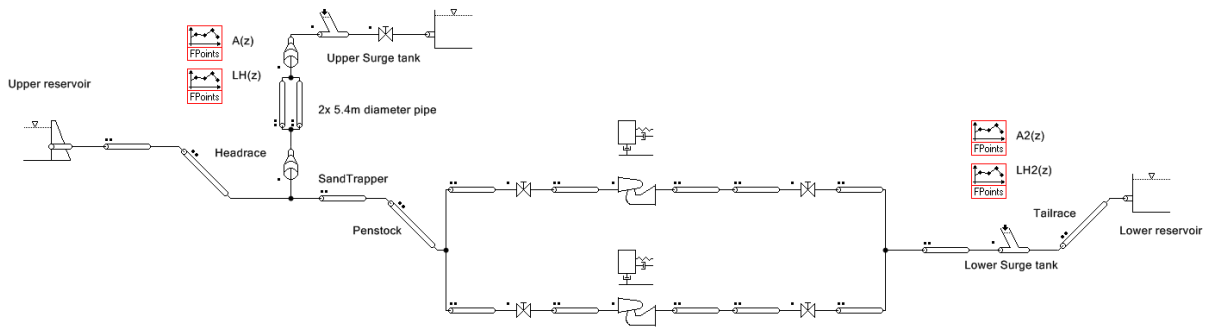

*Figure 8 1D SIMSEN simulation model of the 2 x 395 MW Frades 2 pumped storage power plant.*

Table 2. Frades 2 pump storage power plant characteristics.

| **Francis pump-turbine** | |
|---|---|
| Type | Francis type single-stage reversible pump-turbine |
| Head | Maximum 431.80 m, Minimum 413.64 m |
| Number of units & unit size | 2 units, 4.500 m |
| Turbine rotational speed range | 350 min$^{-1}$, 381 min$^{-1}$ |
| Mechanical power | Generating mode: 400 MW, 390 MW, 190 MW |
|  | Pumping mode: -300 MW, -381 MW, -390 MW |
| Rated mechanical power | 395 MW |
| Specific speed number | 38 SI |
| Mechanical time constant | 7.9 s |
| **Motor-Generator** | |
| Type of power generator | Asynchronous machine |
| Variable speed | DFIM |
| Rated power | 420 MVA |
| Network frequency | 50 Hz |

### 3.2. FCR evaluation

The quantification of the Frequency Containment Reserve capability of the Frades 2 PSP with the two variable speed units is assessed by performing numerical simulations with the simulation conditions and scenarios described in the section 2.3.1. By taking advantage of the so-called flywheel effect, variable speed enables the fast active power injection/absorption in pumping and generating mode. The additional degree of freedom offered by the variable speed opens the door to different control strategies for the hydroelectric plant, as the converter can be used for speed or power control [5]. This can be exploited to maximize either the efficiency or the power reserve dedicated to grid support. The FCR capacity is evaluated with the most obvious approach for exploiting the variable speed unit, corresponding to using the converter for power control while the pump-turbine regulates the speed. Usually, the governor targets the speed that maximizes turbine efficiency for given power. If there is a need for a sudden increase in power supply to the grid, the rotational kinetic energy from the inertia is transferred to the grid, causing a reduction in the speed of the unit, which is then compensated for by the speed controller. In order to increase the FCR capacity, it can thus be advantageous to set the initial speed of the unit in the middle of the allowable range, $n_{middle}$ = 365.5 min$^{-1}$, in order to have the maximum flexibility for under and over speed excursion.

The output power response to a frequency deviation of ±200 mHz and corresponding unit transient behavior with a permanent droop of *Bs=0.85 %*, yielding a FCR active power contribution of ±186.4 MW, is shown in Figure 9. It is noteworthy that, setting the initial speed of the unit at $n_{middle}$ enables to maximize increase the contribution to FCR service to cover the full extended turbine operating range, which is the theoretical maximum possible. It can be noticed that, in the case of a frequency deviation of + 200 mHz, a strategy switch with the motor-generator in speed control is necessary to maintain the unit overspeed within the admissible limits. Nevertheless, the power response of the unit remains compliant with the grid code considered.

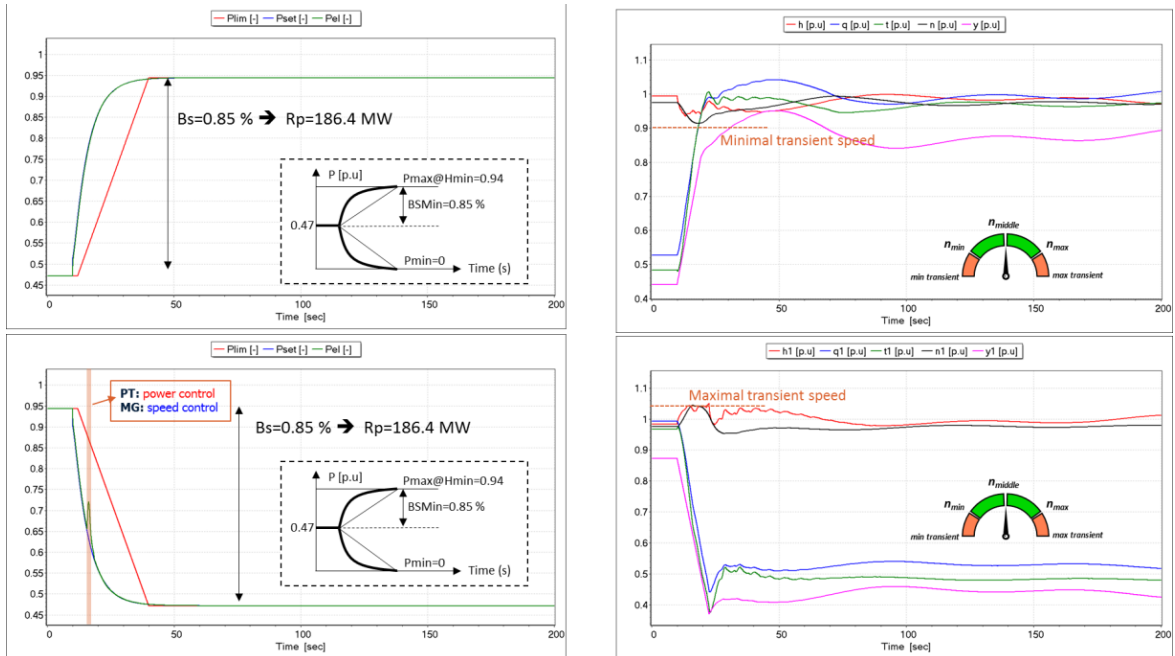

*Figure 9 FCR in turbine mode - Output power response (left) and corresponding unit transient behavior (right) to a frequency deviation of ±200mHz with Bs=0.85% and $n_{ini}=n_{middle}$.(right graphs with h:net head, q: discharge, t: torque, n: rotational speed, y: guide vane opening of unit 1).*

In pump mode, the input power is mainly determined by the rotational speed of the unit. Given the Frades pump operating range comprised between -300 MW and -390 MW, cf. Table 2, and the symmetry of the FCR ancillary service, the maximum power control band is obtained by initially operating the machine in the mid-power and is equal to Rp= ± 45 MW. The resulting input power response to a frequency deviation of ±200 mHz and corresponding unit 1 transient behavior are shown in the Figure 10. Remarkably, the variable speed technology allows the FCR power band to be equal over the entire operating range of the PSP in pump mode without any difficulty.

During hydraulic short circuit mode, the magnitude of the FCR services which can be delivered at the plant level is the sum of all the individual unit services contributing to a given services at the same time, as has been verified by simulation. The total FCR capacity of the Frades 2 PSP in hydraulic short circuit mode is therefore equal to 186.4 MW + 45 MW = 231.4 MW.

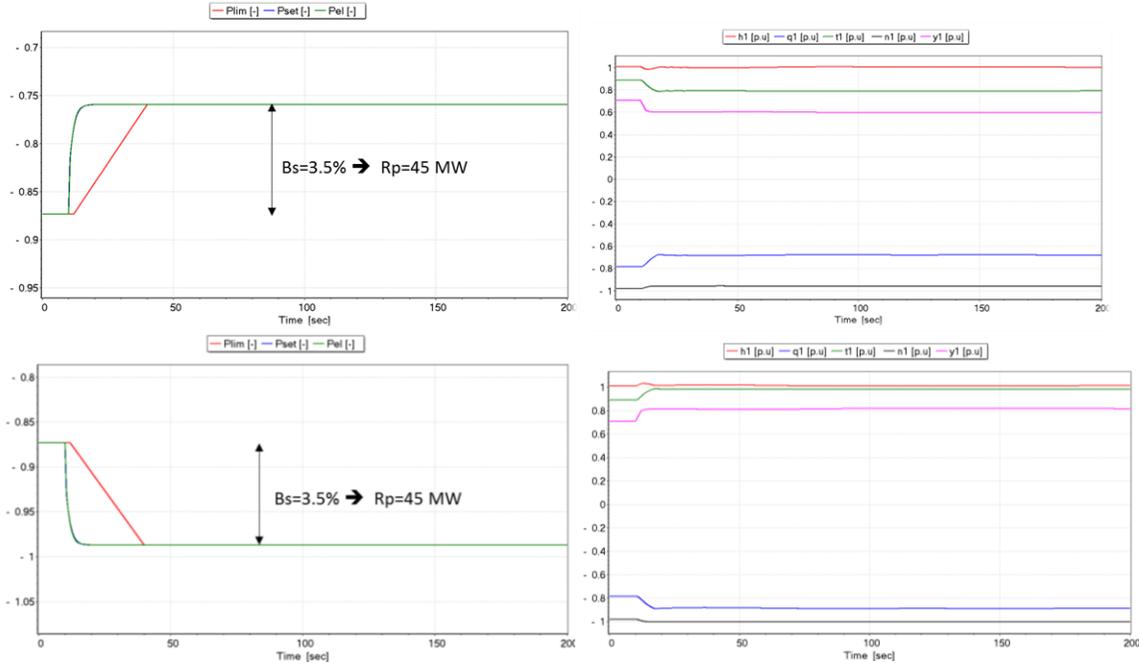

*Figure 10 FCR in pump mode - Output power response (left) and corresponding unit transient behavior (right) to a frequency deviation of ±200mHz with Bs=3.5% and $n_{ini}=n_{middle}$. (right graphs with h:net head, q: discharge, t: torque, n: rotational speed, y: guide vane opening of unit 1).*

### 3.3. aFRR evaluation

The simulations aFRR auxiliary service qualification tests were carried out with both units operated in turbine or pump mode, as well as in HSC mode. This section presents the results obtained in HSC mode, which effectively encompasses the other scenarios. In this scenario, one of the units is operated in turbine mode with an initial power corresponding to the minimum normal operating power of $P_{min}$ = 186.4 MW, while the other unit is in pump mode at the maximum input power of $P_{max}$ = 390 MW. To simulate a power demand on the grid, the turbine power increases from $P_{min}$ to $P_{max}$ in 300 s, while the pump input power is decreased from $P_{max}$ to $P_{min}$ = 300 MW. The resulting power response and unit transient behaviors are shown in the Figure 11. To be compliant with the grid code, the output power indicated by the black curve must lie between the upper and lower limits displayed by the red and blue curves respectively. It is observed that the output power of the unit is able to follow the setpoint almost perfectly. This is because the fast dynamics of the power electronic of the converter and the slow variation of the setpoint allow this slow power ramp to be followed without problems. It is observed that the transient speed of the unit in turbine mode oscillates around the set point, which is due to the flywheel effect. Nevertheless, the turbine speed governor is able to maintain the speed within the allowed range.

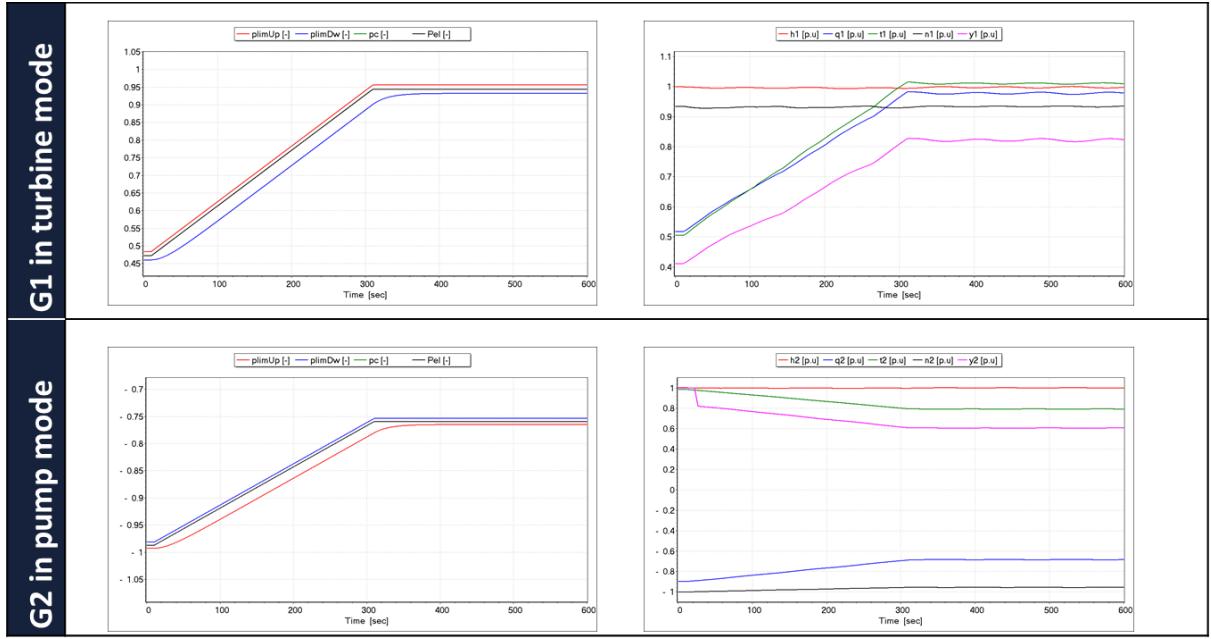

*Figure 11 Variable speed - aFRR in HSC mode - power response (left) and corresponding transient behaviour of unit 1 in turbine mode and unit 2 in pump mode (right) to a grid power increase .(right graphs with h:net head, q: discharge, t: torque, n: rotational speed, y: guide vane opening of unit 1).*

### 3.4. FFR evaluation

To assess the FFR capability, a scenario is simulated with both units of Frades 2 operated in turbine mode under the minimum head and at an initial power equal to $P_{max}$ at $H_{min}$, minus 1.2 times the FFR capacity. When the power converter supplies power to the network by adjusting the electromagnetic torque on the pump-turbine, the unit experiences a decrease in speed, which is then corrected by the speed governor of the pump-turbine at a second time. As the dynamics of the pump-turbine regulator are slower than that of the converter, the speed deviation must however be restricted within the speed limits inherent to DFIM technology. These considerations open the door to different possible strategies of turbine operation for the provision of FFR service. For instance, the turbine can initially be operated at the speed which maximizes the unit efficiency $n_{opt=nmin}$, the middle speed $n_{middle}$, which maximizes the FCR capacity, or the maximum speed $n_{max}$, which maximizes the energy stored in the rotating masses [5].

The results for the maximum FFR capacity assessment in turbine mode related with the initial speed at $n_{middle}$ is shown in the Figure 12. In this configuration, the FFR capacity of each unit in turbine mode is thus equal to 80 MW delivered within 1.3 s, which corresponds to the power step that brings the machine just above the minimum allowable transient speed. It should be mentioned that, in order to optimize the FFR capacity and not to cause a more severe speed drop than necessary, the power setpoint transmitted to the machine is ramped with a slope corresponding to the requirement of the grid code, *i.e.* 1.3 s for 100% of the FFR power supplied. In the same spirit, when the frequency is restored after 30 s, the power setpoint used to bring the unit back to its initial power level features a slow slope, to avoid an overspeed of the unit beyond the admissible value.

Finally, it's worth mentioning that FFR capacity can be increased by using the full range of available speeds. If the energy stored in the rotating masses is maximized by selecting the initial speed $n_{max} = 381$ min$^{-1}$, the FFR capacity of each unit in turbine mode can be equal to 110 MW delivered in 1.3 s [5].

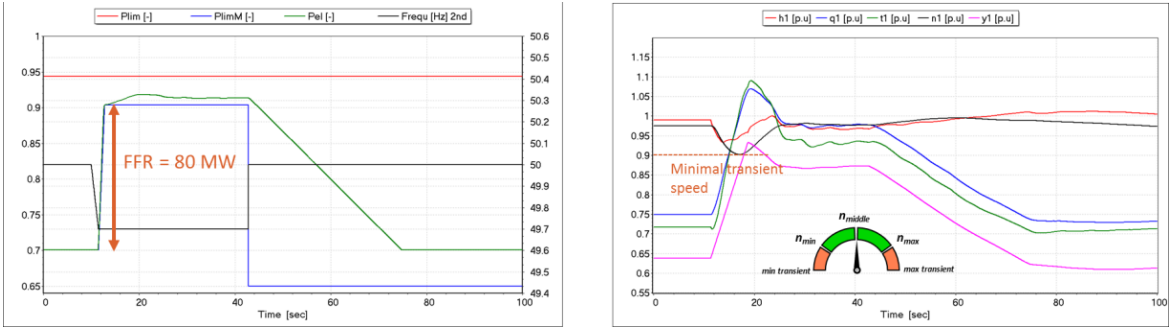

*Figure 12 FFR in turbine mode with initial speed at $n_{middle}$. Output power response (left) and corresponding unit transient behavior (right) .(right graphs with h:net head, q: discharge, t: torque, n: rotational speed, y: guide vane opening of unit 1).*

### 3.5. Black-start evaluation

To evaluate the black-start capability of Frades 2 PSP with variable speed units, simulations are conducted to determine the maximum electrical load that can be connected to the power network, considering that the unit can be operated in so-called grid forming instead of grid-following, [9], [10]. By assuming a purely resistive load connection, this corresponds to simulating a step change in active power, with the unit stabilized at speed no-load condition, and the turbine governor in speed control mode. By exploiting the flexibility of the variable speed, the initial speed at which the unit is stabilized can be chosen arbitrarily within the allowable operating range. For each of the initial conditions, the maximum power step that keeps the unit's transient speed within acceptable limits is sought. Figure 13 illustrates the output power response and corresponding transient behavior of one unit when delivering the maximum black start capacity of 113 MW for the initial rotational speed set at $n_{middle}$. Simulations indicate that the highest black-start capacity can be achieved by operating the unit at a speed of $n_{max}$ = 381 min$^{-1}$ , with an electrical load of 124 MW which can be connected to the unit.

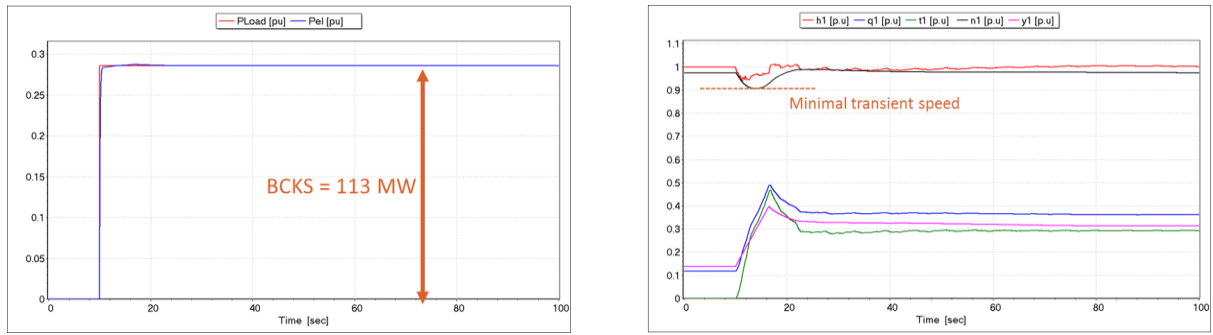

*Figure 13 Variable speed - BCK in turbine mode with initial speed at $n_{middle}$ - Output power response (left) and corresponding transient behaviour of unit 1 (right) .(right graphs with h:net head, q: discharge, t: torque, n: rotational speed, y: guide vane opening of unit 1).*

### 4. Ancillary Services Matrix (ASM)

The ASM aims to map the different flexibility services, associated flexibility markets, as well as the different technologies options to be integrated in the HPPs to endow them with enhanced capacity to provide the flexibility services. Moreover, the development of the XFLEX HYDRO project is allowing to provide an explicit indication of the level of adequacy of each demonstrator, equipped with a new technological solution, to provide flexibility services. Following this rationale, the complete version of the ASM has been elaborated based on more than 130 simulations performed to quantify all ancillary services for the different demonstrators and associated technologies.

The systematic numerical simulation and analytical approaches of ancillary service contribution of each demonstrator and related technologies enabled to produce a scoring system between 0 (no capability) and 5 (full capability) that has been introduced for each ancillary service, allowing to populate with the corresponding scores the Ancillary Service Matrix, presented in Figure 14.

In this matrix, each score refers to the service capability of one unit, not the whole power plant. The magnitude of the services which can be delivered at the plant level is the sum of all the individual units contributing to a given service at the same time. This is also applicable to the HSC operation, where the power plant contribution corresponds to the aggregation of turbine and pump services. In addition, it should be mentioned that all services were evaluated independently, nevertheless power plant operators can provide a combination of some grid services at the same time, provided that the units have sufficient provision from capacity perspectives, and thus it is subject to an economical optimisation from the electricity market perspective.

As far as the Frades 2 demonstrator is concerned, the analysis of the corresponding Ancillary Services Matrix results it could be concluded that, significant increase of ancillary service provision can be made with the variable speed technology as compared to the fixed speed technology. Indeed, the fast dynamics of the power electronics allows the unit to swiftly follow load variations. The following comments can be made in the case of Frades 2 ancillary services evaluations:

- The variable speed enables the FFR ancillary service, which is not adapted for the case of fixed speed as a very fast dynamic is necessary to be able to provide the requested power within the service timeframe.

- Using the synthetic inertia, the variable speed is able to give the same inertia support to the grid as a conventional fixed speed unit as shown in [11],

- With variable speed, the FCR and aFRR ancillary services are enabled over the full operating range in pump mode whereas with fixed speed, the input power is fixed by the rotational speed for a given head, so that provision of the ancillary service is not possible.

- Large improvement of Black-Start capacity can be achieved with variable speed. Especially by choosing a strategy which fully exploits the flywheel effect such as setting the initial rotational speed at $n_{max}$.

- In fixed speed, while the pump power is not modulable individually, the Frades 2 PSP can still provide ancillary service with the power plant in pump mode by operating the turbine simultaneously in hydraulic short circuit.

- During hydraulic short circuit mode, the magnitude of the ancillary services which can be delivered at the plant level is the sum of all the individual unit services contributing to a given services at the same time. That is to say that the ancillary service in pump mode enabled by the variable speed can be cumulated with the corresponding services in turbine mode.

- SPPS allows, in principle, to significantly extend the operating range in turbine mode, while no extension is expected in pump mode due to stability limits at low discharge and cavitation or power limitation at high discharge. The SPPS used in combination with variable speed, enable to have a FCR power band equal to the whole extended operating range of the power plant between 0 MW to $P_{max}$ at Hmin. Whereas with fixed speed technology, the FCR power band remain restricted despite the used of SPPS, due to the hydraulic system limitations.

- Given that for aFRR there are no hydraulic system limitation, the SPPS with fixed and variable speed allows for a aFRR power reserve to be equal to the whole extended operating range in turbine mode, *i.e.*, between 0 MW to Pmax at $H_{min}$.

Similar analysis can be performed to the others demonstrators and associated technologies, enabling to draw the general conclusions and synthesis and summary table provided in the following chapter.

*Figure 14 Ancillary Services Matrix, see [18].*

## 5. Key messages from the ASM

The scoring presented in the ASM of Figure 14 with respect to the different demos/technologies and evaluated ancillary services provides relevant conclusions:

- For Alqueva II and Vogelgrün, **the use of variable speed enables to provide FCR service and overcome the adverse power response of fixed speed technology**, due to water way dynamics.
- **With variables speed technology, the contribution to FFR is strongly influenced by the rotating inertia of the unit in turbine mode** and larger mechanical time constant enable higher FFR contribution; **surprisingly, this is not applicable in pump mode** where the rotating inertia has no influence on the ability to provide FFR service.
- **Synthetic inertia is equivalent to synchronous inertia in terms of ability to inject or absorb active power** in the power network in case of grid frequency variation provided the power network already include sufficient synchronous inertia.
- **Variable speed technology enables to increase black-start capability by a factor ranging between 2 and 10 as compared to fixed speed technology**, provided that the variable speed control encompasses grid forming capability.
- **HSC enables fixed speed power plants to provide FCR and aFRR services as well as adjustable mFRR and RR services, when operating in pump mode**, with a power range corresponding to the turbine power range.
- **SPPS enables to take advantage of the full turbine power range between 0 and 100%** of rated power to provide aFRR, mFRR and RR services with typical extension from 40% to 100% of the power range.
- **HBH enables the Vogelgrün demonstrator to overcome the adverse power response** due to waterway dynamics and **to contribute to FCR** while reducing the associated wear and tear.

Since the results presented in the ASM are rather dense, a summary table of ancillary services was elaborated and is presented in Figure 15 in order to synthetise the key results of the Ancillary Services Matrix, presented by type of technology and grid services.

*Figure 15 Summary table of ancillary services.*

# 6. Conclusions

The systematic 1D SIMSEN numerical simulation of ancillary service contribution of each demonstrator and related technologies performed within the Work Package 2 of the XFLEX HYDRO European Project enabled to quantify the magnitude of active power response to contribute to FFR, FCR, aFRR and Black-Start. Since FCR and aFRR are bidirectional services with power increase or decrease, overall more than 130 simulations have been performed and documented to quantify all ancillary services. The results expressed first in MW and per unit have been then converted into a scoring between 0 and 5 which was introduced for each ancillary service, allowing to populate the so-called Ancillary Service Matrix presented in Figure 14. Moreover, synchronous inertia, synthetic inertia, mFRR, RR and Voltage/Var services have also been scored based on simplified approaches and the corresponding scores have also been included in the Ancillary Services Matrix. The analysis of the score of the Ancillary Services Matrix enable to draw several conclusions available in the chapter 4. The key conclusions obtained for the different technologies addressed within XFLEX HYDRO project have been then synthetised in chapter 5 and summarised in a table presented in Figure 15 which gives a global overview of possible improvements of ancillary services that the implementation of a given technology can bring.

The time series of simulation results obtained for the FFR, FCR and aFRR ancillary services have also been used to perform parameter identification of simplified hydropower dynamic models, necessary to perform large-scale hydropower flexibility assessment. Dynamic simulations have been carried out across a simplified European grid model, to assess the behaviour of the European interconnected system under the deployment extended hydropower flexibility solutions proposed in the project [19].

Finally, the Ancillary Service Matrix presented in Figure 14 was made available publicly on the XFLEX HYDRO website as a PDF document, [18]. This document will be a powerful communication tool to promote and highlight the flexibility gain and the related benefits that the different technologies addressed in the XFLEX HYDRO can bring to hydroelectric power plant operators. This matrix will be of high significance for the hydro industry, generation companies and policy makers since it will provide a resumed mapping of the hydro technology regarding the provision of flexibility services and the capability of participating in new power markets. Ultimately, all the results produced within the Work Package 2 will also contribute to the technical White Paper which will be made public at the end of the project.

# 7. Annex 1: Short overview of the XFLEX HYDRO demonstrators

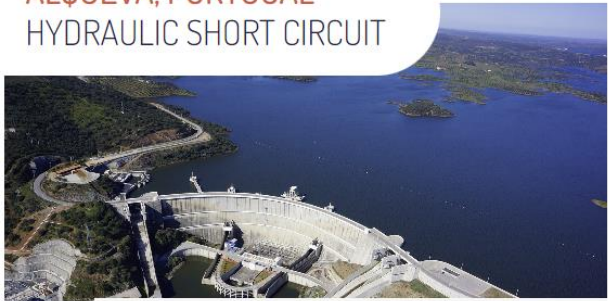
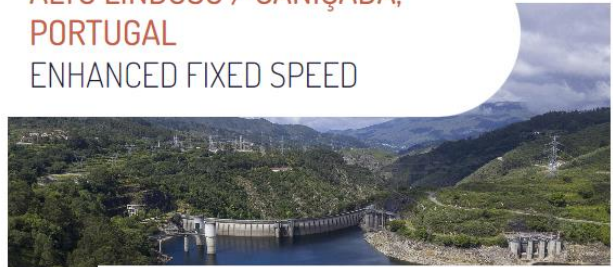
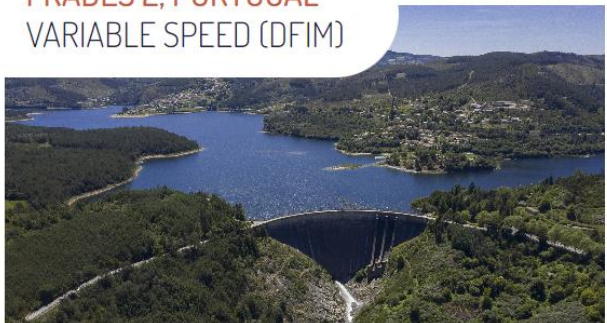
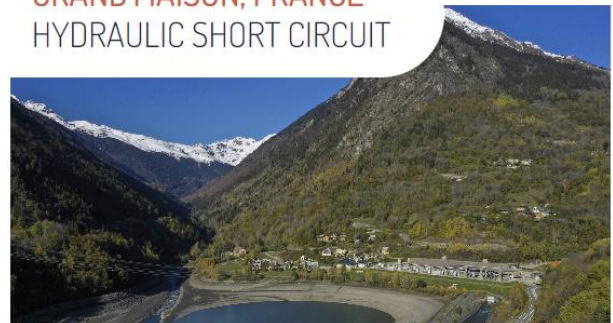
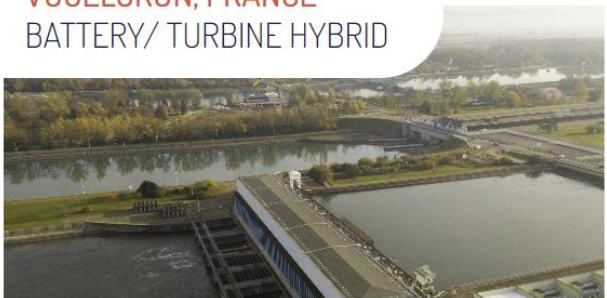
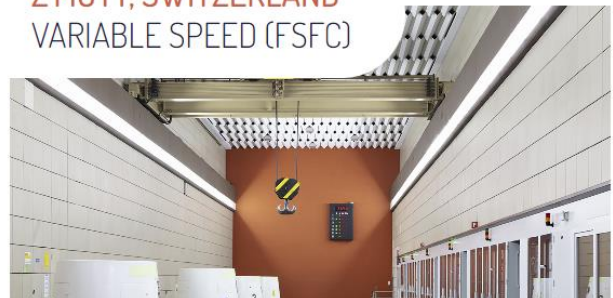

*Figure 16 Short overview of the XFLEX HYDRO demonstrators, for more information see [1].*

## 8. Acknowledgements

The Hydropower Extending Power System Flexibility (XFLEX HYDRO) project has received funding from the European Union's Horizon 2020 research and innovation programme under grant agreement No 857832.

**The Authors**

**Christophe Nicolet** graduated from the Ecole polytechnique fédérale de Lausanne, EPFL, in Switzerland, and received his Master degree in Mechanical Engineering in 2001. He obtained his PhD in 2007 from the same institution in the Laboratory for Hydraulic Machines. Since, he is managing director and principal consultant of Power Vision Engineering Sàrl in St-Sulpice, Switzerland, a company active in the field of optimization of hydropower transients and operation. He is also external lecturer at EPFL in the field of "Transient Flow".

**Matthieu Dreyer** graduated from the Ecole polytechnique fédérale de Lausanne, EPFL, in Switzerland, and received his Master degree in Mechanical Engineering in 2010. He obtained his PhD in 2015 from the same institution in the Laboratory for Hydraulic Machines. After two years of post-doctoral research in the same laboratory, he joined Power Vision Engineering Sàrl in St-Sulpice, Switzerland.

**Christian Landry** graduated from the Ecole polytechnique fédérale de Lausanne, EPFL, in Switzerland, and received his Master degree in Mechanical Engineering in 2010. He obtained his PhD in 2015 from the same institution in the Laboratory for Hydraulic Machines. Since 2016, he is working for Power Engineering Sàrl in St-Sulpice, Switzerland, on transient phenomena, simulation and analysis of the dynamic behavior of hydroelectric power plants and their interactions with the power network.

**Sébastien Alligné** is a project engineer at Power Vision Engineering Sàrl in St-Sulpice, Switzerland. He graduated in Mechanical Engineering at the Ecole Nationale Supérieurs d'Hydraulique et de Mécanique de Grenoble, ENSHMG, in 2002. He then worked in automotive and hydraulic industries between 2002 and 2007. He obtained his PhD degree at EPFL Laboratory for Hydraulic Machines in the field of simulation of complex 3D flow in hydraulic machinery in 2011. Since 2011, he is responsible of hydraulic system dynamic behaviour modelling and of CFD computations at Power Vision Engineering.

**Antoine Béguin** is a project engineer at Power Vision Engineering Sàrl in St-Sulpice, Switzerland. He graduated from the Ecole polytechnique fédérale de Lausanne, EPFL, in Switzerland, and received his Master degree in electrical engineering in 2006. He obtained his PhD in 2011 from the same institution in the Laboratory for Power Electronics. Since 2011, he is working with Power Vision Engineering Sàrl in St-Sulpice, Switzerland, on transient phenomena, simulation and analysis of the dynamic behavior of hydroelectric power plants and their interactions with the power network. As a software development and communication specialist, he is responsible for the development of the Hydro-Clone system.

**Matteo Bianciotto** holds an MSc in Mechanical Engineering from the Polytechnic University of Turin and a Master in Energy Management from the ESCP Europe Business School. He worked as a R&D engineer for Andritz Hydro and at the Laboratory of Hydraulic Machines of the EPFL. Over this period, he oversaw various development projects and cavitation visualisation on Pelton turbines as well as reduced model and onsite thermodynamic tests. Between 2014 and 2021, he worked as a financial consultant in London where he developed advanced economic models for transactions involving hydropower and other energy assets. He is currently focusing on the preparation of a series of industry guidelines and policy recommendations resulting from the XFLEX HYDRO project.

**Richard Taylor** graduated in Earth Sciences and Water Resources from the University of London in 1985. He holds additional post-graduate qualifications in Hydropower Engineering and Sustainable Development. He has worked in the Hydropower sector since 1986, holding leadership positions in both private and not-for-profit organizations. Between 2000 and 2019 he was the Chief Executive Officer of the International Hydropower Association (IHA). Since then, he has worked as a consultant, acting as a Senior Hydropower Expert for the World Bank and the Asian Development Bank, as well as continuing to support the work of IHA. This has included specific deliverables for XFLEX HYDRO throughout the project.

**Manuel Vaz Castro** graduated in Electrical Engineering at the Faculty of Engineering of the University of Porto (FEUP) and received his Master degree in July 2015. From August 2015 to November 2016, he worked as a Power Cable Engineer at Cabelte group. Since February 2017, he is a Power System Researcher at Centre for Power and Energy Systems of INESC TEC. His work has been focused on multi-microgrids operation and control and frequency stability of electric power systems with increasing shares of inverter based resources.

**Maria Helena Vasconcelos** graduated in Electrical Engineering at the Faculty of Engineering of the University of Porto (FEUP) in 1996, completed the Master degree in November 1999 and the PhD in January 2008, all the degrees in Power Systems from FEUP. She is a Researcher in the Centre for Power and Energy Systems of INESC TEC since September 1996, and a Senior Researcher since January 2008. In November 2002 she joined the Department of Electrical Engineering of FEUP as an Invited Assistant Professor and, as an Assistant Professor, since January 2008. There, she lectures several classes in graduation, MSc and PhD courses in Electrical Engineering and Power Systems.

**Carlos Moreira** graduated in Electrical Engineering in the Faculty of Engineering of the University of Porto - FEUP (2003) and completed his PhD in Power Systems in November 2008, also from the University of Porto. He is a Senior Researcher in the Centre for Power and Energy Systems of INESC TEC since September 2003. In February 2009 he joined the Department of Electrical

Engineering of FEUP as Assistant Professor. He lectures several classes in graduation and MSc courses in Electrical Engineering and Power Systems and supervises the research activities of several MSc and PhD students. His main research interests are related to microgrids operation and control, dynamics and stability analysis of electric power systems with increasing shares of converter interfaced generation systems and grid code development.